\documentclass[12pt]{article}
\usepackage[english]{babel}
\usepackage{epsfig}
\begin{document}
\large

\par
\noindent {\bf Cherenkov effect in the weak interactions generated
by the neutrinos and new approach for estimation of neutrino mass}
\par
\begin{center}
\vspace{0.3cm} Beshtoev Kh. M. (beshtoev@cv.jinr.ru)
\par
\vspace{0.3cm} Joint Institute for Nuclear Research, Joliot Curie
6, 141980 Dubna, Moscow region, Russia.
\end{center}
\vspace{0.3cm}

\par
Abstract

\par
It is shown that if weak interactions can generate masses and
polarize matter, then the Cherenkov effect induced by these
interactions appears. The resonance ($v_\nu < c/n$) and the
Cherenkov ($v_\nu > c/n$) effects are competitive processes and at
definite neutrino energies the resonance effect will change to the
Cherenkov effect and we obtain an excellent possibility of
estimating neutrino masses.

\section{Introduction}

At present the existence of tree types of neutrinos- electron
($\nu_e$), muon ($\nu_\mu$) and tau ($\nu_\tau$) neutrinos - is
established \cite{1}. Determination of masses of these neutrinos
is of great interest. Experiments were carried out to estimate the
electron neutrino mass with using the beta decay \cite{2}. Also,
experiments on neutrinoless double beta decays were conducted to
estimate neutrino masses on the assumption that neutrinos are
Majorana particles \cite{3}. In addition, mass differences between
$\nu_1, \nu_2, \nu_3$ were measured in the neutrino oscillation
experiments \cite{4}, but in these experiments it is impossible to
establish neutrino masses.
\par
The suggestion that, by analogy with $K^{o},\bar K^{o}$
oscillations, there could be neutrino-antineutrino oscillations (
$\nu \rightarrow \bar \nu$), was considered by Pontecorvo \cite{5}
in 1957. It was subsequently considered by Maki et al. \cite{6}
and Pontecorvo \cite{7} that there could be mixings (and
oscillations) of neutrinos of different flavors (i.e., $\nu_{e}
\rightarrow \nu_{\mu}$ transitions). Then the resonance mechanism
of neutrino oscillations in matter \cite{8} was assumed which
implied that as neutrinos passed though matter, enhancement of
neutrino oscillations took place since effective masses of
neutrinos change and at definite matter density they can be equal.
\par
This work is devoted to consideration of Cherenkov radiation of
neutrinos in matter and a new approach for estimation of the
neutrino mass.
\par
Cherenkov radiation can appear only when neutrinos move in matter
with a velocity $v_i > c/n_i, \quad i = \nu_e, \nu_\mu, \nu_\tau$.
But at the neutrino velocity $v_i < c/n_i$ there may take place
resonance enhancement of neutrino oscillations in matter.
Therefore, before considering the Cherenkov effect, we give
elements of the resonance effect.

\section{Elements of the resonance mechanism enhancement of neutrino
oscillations in matter}

\par
Before consideration of the resonance mechanism it is necessary to
gain an understanding of the physical nature of origin of this
mechanism. As neutrinos pass through matter, there can be two
processes: neutrino scattering and polarization of the matter by
neutrinos. Obviously, resonance enhancement of neutrino
oscillations in matter will arise due to polarization of the
matter by neutrinos. If the weak interaction can generate not only
neutrino scattering but also polarization of matter, then the
resonance effect will exist otherwise this effect cannot exist.
\par
In the ultrarelativistic limit, the evolution equation for the
neutrino wave function $\nu_{\Phi} $  in matter has the following
form \cite{8}:
$$
i \frac{d\nu_{Ph}}{dt} = ( p\hat I + \frac{ {\hat M}^2}{2p} + \hat
W ) \nu_{Ph} , \eqno(1)
$$
where $p, \hat M^{2}, \hat W_i $ are, respectively, the momentum,
the (nondiagonal) square mass matrix in vacuum, and  the matrix,
taking  into account neutrino interactions in matter,
$$
\nu_{Ph} = \left (\begin{array}{c} \nu_{e}\\
\nu_{\mu} \end{array} \right) , \qquad \hat I = \left(
\begin{array}{cc} 1&0\\0&1 \end{array} \right) ,
$$
$$
\hat M^{2} = \left( \begin{array}{cc} m^{2}_{\nu_{e}\nu_{e}}&
m^{2}_{\nu_{e} \nu_{\mu}}\\ m^{2}_{\nu_{\mu}\nu_{e}}&
m^{2}_{\nu_{\mu} \nu_{\mu}} \end{array} \right).
$$
\par
If we suppose that neutrinos in matter like photons in matter
(i.e., the polarization at neutrino passing through matter arises)
and the neutrino refraction indices are defined by the expression
$$
n_{i} = 1 + \frac{2 \pi N}{p^{2}} f_{i}(0) = 1 + 2 \frac{\pi
W_i}{p} , \eqno(2)
$$
where $i$ is a type of neutrinos $(\nu_e, \nu_\mu, \nu_\tau)$, $N$
is density of matter, $f_{i}(0)$ is a real part of the forward
scattering amplitude which appears owing to polarization of matter
by neutrino, then $W_i$ characterizes polarization of matter by
neutrinos (i.e. it is the energy of matter polarization).
\par
The electron neutrino ($\nu_{e}$)  in matter interacts via
$W^{\pm}, Z^{0}$ bosons and $\nu_{\mu}, \nu_{\tau}$ interact only
via the $Z^{0}$ boson. These differences in interactions lead to
the following differences in the refraction coefficients of
$\nu_{e}$ and $\nu_{\mu}, \nu_{\tau}$
$$
\Delta n = \frac{2 \pi N}{p^{2}} \Delta f(0) , \eqno(3)
$$
$$
\Delta f(0) =  \sqrt{2} \frac{G_F}{2 \pi} p ,
$$
$$
E_{eff} = \sqrt{p^2 + m^2} + <e \nu|H_{eff}|e \nu> \approx p
+\frac{m^2}{2 p} + \sqrt{2} G_F N_e
$$

where $G_F$ is the Fermi constant.
\par
Therefore the velocities (or effective masses) of $\nu_{e}$ and
$\nu_{\mu}, \nu_{\tau}$ in matter are different. And at the
suitable density of matter this difference can lead to resonance
enhancement of neutrino oscillations in matter \cite{8}, \cite{9}
\par
$$
\sin ^{2} 2\theta _{m} = \sin^{2} 2\theta \cdot [(\cos 2\theta  -
{L_{0}\over L^{0}})^{2} + \sin ^{2} 2\theta ]^{-1} , \eqno(4)
$$
where $\sin ^{2} 2\theta _{m}$ and $\sin^{2} 2\theta$ characterize
neutrino mixing in matter and vacuum, $L_{0}$ and $L^{0}$ are
lengths of oscillations in vacuum and matter
$$
L_{0} = \frac{4 \pi E_{\nu} \hbar}{\Delta m^2 c^3} \qquad L^{0} =
\frac{\sqrt{2} \pi \hbar c}{G_F n_e}, \eqno(5)
$$
where $E_{\nu}$ is the neutrino energy, $\Delta m^2$ is the
difference between squared neutrino masses, $c$ is the velocity of
light, $\hbar$ is the Plank constant, $G_F$ is the Fermi constant
and $n_e$ is the electron density of matter.
\par
\noindent At resonance

$$ \cos  2\theta  \cong  {L_{0}\over
L^{0}}\qquad sin^{2} 2\theta_{m} \cong 1\qquad \theta_{m} \cong
{\pi \over 4} . \eqno(6)
$$
\par
It is necessary to stress that this resonance enhancement of
neutrino oscillations in matter is realized when, the neutrino
velocity is smaller than the velocity of light in matter (i.e. $
v_i < \frac{c}{n_i})$.
\par
What will happen when the neutrino velocity is larger than the
velocity of light in matter?  Now let'us turn to consideration of
this problem.

\section{Cherenkov effect in weak interactions generated by
neutrinos in matter}

The specific electromagnetic radiation produced by fast electrons
moving in a medium was observed by Cherenkov in 1934 \cite{10}.
Tamm and Frank \cite{11} showed that the charged particle must
radiate when its velocity exceeds the velocity of light in the
medium (see also \cite{12} where motion of a charged particle in a
medium with a constant electric permittivity (or refraction index)
was considered). It is obvious that analogous radiation must take
place when neutrinos move in a medium with a velocity exceeding
the velocity of light in the medium if $n_i - 1 > 0$.
\par
For realization of the mechanism of resonance enhancement of
neutrino oscillations in matter the following condition must be
fulfilled:
$$
n_i -1 > 0 , \quad n_{\nu_e}^{SUN} -1 =\Delta n_{\nu_e} \approx
10^{-17} \div 10^{-19} \eqno(7)
$$
which is equivalent to the demand that the matter polarization
exist at
$$
v_i < \frac{c}{n_i} . \eqno(8)
$$
\par
For existence of the Cherenkov effect in weak interactions it is
necessary to fulfil the following two conditions:
$$
n_i -1 > 0 , \eqno(9)
$$
and
$$
v_i > \frac{c}{n_i} , \eqno(10)
$$
where $v_i, c$ are the velocities of the neutrino and light
respectively. The first condition coincides with the resonance
existence condition. The second condition means that if the
neutrino moves with the velocity which is larger than the velocity
of light in matter, then it will polarize matter and go away
keeping in reserve this polarization which must be radiated
afterwards. What will happen after that? This polarization can be
taken off in two ways:
\par
\noindent 1. Via radiation of weakly interacting particles
(neutrinos) or
\par
\noindent 2. Via electromagnetic radiation (polarized electrons
have electrical charges and they can give electromagnetic
radiation).
\par
\noindent Since the probability of radiation of weakly interacting
particles is very small, the Cherenkov radiation will be realized
mainly in the form of electromagnetic radiation but not weak
interaction radiation (it is clear that if energy of matter
polarization is very small then neutrinos cannot be produced). It
is related with the fact that weak interactions are slow processes
while electromagnetic processes are fast. Energy of this radiation
is
$$
E \approx W = \sqrt{2} G_F N_e ,\quad (E^{SUN} \approx
10^{-10}\div 10^{-11} eV) \eqno(11)
$$
where $G_F$ is the Fermi constant, $N_e$ is the electron density
in the matter. The radiation angle $\beta_i$ is
$$
cos \beta_i = \frac{c}{v_i n_i}
$$
\par
If
$$
v_i < \frac{c}{n_i} , \eqno(12)
$$
then the neutrino will polarize matter and since the velocity is
smaller than the velocity of light in matter, this matter
polarization will move together with this neutrino and the
resonance effect can be realized.
\par
Now we see that the resonance and Cherenkov effects are competing
processes. If $ v_i < \frac{c}{n_i}$, the resonance effect will be
realized and if $ v_i > \frac{c}{n_i}$, the Cherenkov effect will
be realized. It is very important to remark that if in reality in
weak interactions the matter polarization is present, then we
obtain an excellent possibility of estimating neutrino masses
using the point transition between the above two mechanisms and
then the expression for the neutrino mass is
$$
m_\nu = E_{trans}{\sqrt{1 -(\frac{1}{n_i})^2}}
$$
if $(n - 1) \ll 1$, then
$$
m_\nu \simeq E_{trans} \sqrt{2 (n_i - 1)} , \eqno(13)
$$
where $E_{trans}$ is the neutrino energy at the point where the
transition between the indicated mechanisms is realized.

\section{Conclusion}

\par
It has been shown that if weak interactions can generate masses
and polarize matter, the Cherenkov effect appears which is induced
by these interactions. The resonance ($v_\nu < c/n_i$) and the
Cherenkov ($v_\nu \ge c/n_i$) effects are competitive processes
and at definite neutrino energies the resonance effect will change
to the Cherenkov effect and we obtain an excellent possibility of
estimating the neutrino masses.
\par
It is necessary to remark that if the mechanism of MaVaN
oscillations \cite{13} is realized then the Cherenkov effect
generated by neutrinos will also be realized.
\\

\end{document}